\documentstyle[preprint,revtex,eqsecnum]{aps}

\draft
\tightenlines
\begin{document}

\preprint{TUM--T31--39/93}
\begin{title}
\begin{center}
\bf THE
 SEMILEPTONIC DECAYS $D\to \pi(\rho) e \nu$ AND $B\to \pi
(\rho) e \nu$ FROM QCD SUM RULES
\end{center}
\end{title}
\author{Patricia Ball}
\begin{instit}
Physik-Department, TU M\"unchen, D-8046 Garching, FRG
\end{instit}
\receipt{\today}
\begin{abstract}
We investigate the semileptonic decays of B and D mesons into $\pi$
and
$\rho$ mesons, respectively, by means of QCD sum rules. We find that
 for the
vector formfactors involved the pole dominance hypothesis is valid to
good accuracy with pole masses in the expected range. Pole dominance,
however, does not apply to the axial formfactors which results in
specific predictions for the predominant polarization of the $\rho$
meson and the shape of the lepton spectrum. For the total decay rates
we find $\Gamma (\bar B^0 \to \pi^+ e^- \bar\nu) = (5.1\pm
1.1)\,|V_{ub}|^2\, 10^{12}\,{\rm s^{-1}}$, $\Gamma ( D^0 \to \pi^-
e^+
\nu) = (8.0\pm 1.7)\,|V_{cd}|^2\, 10^{10}\,{\rm s^{-1}}$, $\Gamma
(\bar
B^0 \to \rho^+ e^- \bar\nu) = (1.2\pm 0.4\,)\,|V_{ub}|^2\,
10^{13}\,{\rm s^{-1}}$ and $\Gamma (D^0 \to \rho^- e^+\nu) = (2.4\pm
0.7)\,|V_{cd}|^2\, 10^{9}\,{\rm s^{-1}}$.
\end{abstract}

\section{Introduction}

Semileptonic weak decays of heavy mesons have proved to be a very
important
tool in exploring the Higgs-sector of the standard model and, in
particular, the strength of weak decays of quarks, parametrized by
the CKM matrix. Once the relevant hadronic matrix elements are known,
the CKM matrix can be extracted from experimental measurements of the
decay rates. Whereas the situation (both experimental and
theoretical)
now seems rather settled for the
dominant decays $B\to D^{(*)} e \nu$ and $D\to K e \nu$, at
least at a level of accuracy of $\sim$~10\% for the CKM matrix
elements \cite{stone,bstone}, the experimental results for the
Cabibbo-suppressed decays $B,\ D\to\pi, \rho e \nu$ at present still
suffer from large statistical uncertainties. This situation, however,
 will
improve in the near future, since the exploration of these decays
is motivated by the quest for $|V_{ub}|$ which of all the CKM matrix
elements still is the one most poorly known
\cite{ARGUS,CLEOinklu,CLEO}.

There exist several theoretical calculations employing
relativistic \cite{BWS,KS} or non-relativistic \cite{GISW} quark
models as well as (for D decays) lattice calculations \cite{LMS,BKS},
and, quite recently, attempts to relate formfactors of D decays to
those of the corresponding B decays by means of the heavy quark
effective theory \cite{gatto}. All these models are,
however, for conceptional or, as for lattice calculations, for
economical reasons not capable of calculating the full dynamics of
the
decay process, even apart from model limitations. A quite standard
procedure is to determine a formfactor $f$ at some fixed point of
$t$,
the momentum transfer squared to the leptons, and then to assume
either some
pole-like $t$-dependence,
\begin{equation}\label{eq:poldominanz}
f(t) \sim \frac{1}{m_{pol}^2 - t},
\end{equation}
where $m_{pol}$ is the mass of the lowest lying resonance coupling to
the corresponding current (\cite{BWS}, e.g.), or an exponentially
increasing formfactor as in the non-relativistic
model of \cite{GISW}. Indeed, at the level of a desired accuracy
of, say, $\sim 20$\% for the rates, the details of the functional
dependence do not matter as long as $t_{max}$, the maximum
value of $t$ allowed by kinematics, is much smaller than $m_{pol}^2$
and the formfactors vary only slowly (as for $B\to D^{(*)} e \nu$).
A certain deviation from that insensitivity is noticeable in the
decay
$D\to K^* e \nu$ where all ``conventional'' (quark model and
lattice)
calculations are not capable to reproduce neither the absolute value
of the rate nor the small value of the ratio of rates of
longitudinal
to transversal polarized $K^*$ (cf.\ \cite{stone}). The actual
functional form of the $t$-dependence becomes crucial if
$t^{max}/m_{pol}^2\approx 1$ like in the decays $B\to \pi,\ \rho e
\nu$
which in the future will provide us with the most accurate
information
on $V_{ub}$. Thus a point seems to be reached where an increased
attention should be paid to the investigation of the $t$-dependence
of
formfactors.

In fact, there is another method for calculating hadronic
matrix-elements
including non-perturbative effects which relies on the
field-theoretical aspects and features of QCD and was designed to
make
maximum use of known manifestations of non-perturbative QCD, the QCD
sum rules method \cite{SVZ}. Originally invented for the
calculation of vacuum-to-meson transition amplitudes, it soon found
application to the calculation of the electromagnetic formfactor of
the pion \cite{smilga} and other meson-to-meson transition amplitudes
(cf.\ \cite{narr} for a review). Although this method in general
yields less detailed
results than fine-tuned models, it has got the advantage that
only a small number of parameters is needed that have an evident
physical meaning (e.g.\ quark masses) and/or characterize the
non-perturbative regime of QCD (e.g.\ the so-called quark
condensate, the order parameter of chiral symmetry breaking). Once
these parameters are fixed from well known processes,
 they can be used to calculate for instance heavy meson decays.

In previous publications \cite{BBD,BPi,ball,BD}, we have shown that
the
$t$-dependence of the formfactors of $D\to K^{(*)}$ and $B\to
D^{(*)}$
can reliably be calculated by means of QCD sum rules. As a general
pattern to be followed, formfactors determined by vector currents
observe a pole-type behaviour with pole masses in the expected
range,
those determined by axialvector currents in general do not. In view
of the above stated advantages of the QCD sum rules method, we
feel it worthwile to apply it likewise to the Cabibbo-suppressed
decays $B,\ D\to \pi,\ \rho e \nu$ and to go beyond the existing QCD
sum rule calculations
\cite{unpub,dompfaffD,ovchi,slobo,narrdecay,dompfaffB} which for
reasons to be explained in the next section were restricted to a
determination of the formfactors at $t=0$. We expect to gain a
reliable
picture of the dynamics of these decays and well founded predictions
of their decay rates within the scope of accuracy to be obtained by
QCD sum rules, i.e.\ at the level of 20\% at best.

Our paper is organized as follows: in Sec.~II we present the QCD sum
rules method, improve the existing calculations and collect all
necessary kinematics. In Sec.~III we evaluate the sum rules and give
results for the decays $D\to \pi,\ \rho e \nu$ and $B\to \pi,\ \rho e
\nu$. Finally, in Sec.~IV we discuss the results and compare it both
to experiment and to other calculations. Formul\ae\ and technical
details are collected in the appendices.

\section{The Method}

We consider the three-point functions
\begin{eqnarray}
\Pi^{\mu\nu} & = & i^2\int d^4x\, d^4y\, e^{-ip_H x + i (p_H-p_L) y}
\, \langle 0 | T j_L^{A\nu}(0) V_{hl}^\mu(y) j_H^\dagger (x) | 0
\rangle\nonumber\\
& = & i (p_H+p_L)^\mu p_L^\nu \Pi_+ + \dots\label{eq:corrpi}
\end{eqnarray}
and
\begin{eqnarray}
\Gamma^{\mu\nu} & = & i^2\int d^4x\, d^4y\, e^{-ip_H x + i (p_H-p_L)
 y}
\, \langle 0 | T j_L^\nu(0) (V_{hl}-A_{hl})^\mu(y) j_H^\dagger (x) |
 0
\rangle\nonumber\\
& = & i g^{\mu\nu} \Gamma_0 - i (p_H+p_L)^\mu p_H^\nu \Gamma_+ -
\epsilon^{\mu\nu}_{\phantom{\mu\nu}\rho\sigma} p_H^\rho p_L^\sigma
\Gamma_V + \dots,\label{eq:corrga}
\end{eqnarray}
respectively. Here $(V_{hl}-A_{hl})_\mu = \bar l \gamma_\mu (1-
\gamma_5) h$ is
the weak current mediating the weak decay of the heavy quark $h$ with
mass $m_h$ into the light quark $l$ with mass $m_l$, $j_H = \bar q
i\gamma_5 h$ is an interpolating field describing the pseudoscalar
meson $H$ ($B$ or $D$) built up from $h$ and the light antiquark
$\bar
q$, and $j_L^{(A)\nu} = \bar q \gamma^\nu(\gamma_5) l$ interpolates
the light vector (pseudoscalar) meson $L$ ($\pi$ or $\rho$). $p_H$
and
$p_L$ are the momenta of the heavy and the light meson, respectively.
In the above equations, we have made explicit only those
Lorentz-structures that actually contribute to the decays under
consideration.

The correlation functions (\ref{eq:corrpi}) and (\ref{eq:corrga}) are
functions of the scalars $p_H^2$, $p_L^2$, and $t=(p_H-p_L)^2$ and
can
be calculated in perturbation theory for Euclidean values
$p_H^2-m_h^2$,
$p_L^2-m_l^2 \ll 0$. On the other hand, the singularity structure of
the correlation functions is known, and thus we
can represent them by double dispersion relations in $p_H^2$ and
$p_L^2$, e.g.
\begin{equation}\label{eq:linkeseite}
\Pi_+ = \int ds_H\, ds_L\,\frac{\rho^{\rm\scriptsize phys}_+
(s_H,s_L,t)}{(s_H-p_H^2)(s_L-p_L^2)} + {\rm subtractions}.
\end{equation}
The spectral function $\rho^{\rm\scriptsize phys}_+$ can be
expressed in terms of physical observables as
\begin{eqnarray}
\rho^{\rm\scriptsize phys}_+ & \sim & (2\pi)^6 \sum_{m,n} \int
\prod_{i,j}^{m,n} \left[ \frac{d^3p_{Li}}{(2\pi)^3 2 E_{Li}}
\frac{d^3p_{Hj}}{(2\pi)^3 2 E_{Hj}} \right]
\delta^4(q_L-{\textstyle\sum} p_{Li}) \delta^4(q_H - {\textstyle\sum}
p_{Hj})\nonumber\\
& & \langle 0 | j_L^{A\nu} | m \rangle \langle m | V_{hl}^\mu | n
\rangle \langle n | j_H^\dagger | 0 \rangle
\end{eqnarray}
where one has to take the appropriate Lorentz-structure on the right
hand side and $q_H^2 = s_H$ and $q_L^2 = s_L$. The sum runs over all
 $m$- and
$n$-particle states coupling to the currents $j_L^{A\nu}$ and $j_H$,
respectively. In particular, we shall single out the ground states
and
write
\begin{equation}\label{eq:rhocont}
\rho^{\rm\scriptsize phys}_+ \sim \langle 0 | j_L^{A\nu} | L \rangle
 \langle
L | V_{hl}^\mu | H \rangle \langle H | j_H^\dagger | 0 \rangle +
\rho^{\rm\scriptsize cont},
\end{equation}
where $\rho^{\rm\scriptsize cont}$ contains both the contributions of
higher resonances with appropriate quantum numbers and  of
many-particle states. The first term on the right
hand side contains exactly the quantities we are interested in:
\begin{eqnarray}
\langle \pi | V_{hl}^\mu | H \rangle & = & f_+(t) (p_H+p_\pi)_\mu +
 f_-(t)
(p_H-p_\pi)_\mu,\\
\langle \rho,\lambda | V_{hl}^\mu-A_{hl}^\mu | H \rangle & = &
-i (m_H + m_\rho) A_1(t) \epsilon_\mu^{*(\lambda)} +
\frac{iA_2(t)}{m_H
+ m_\rho} (\epsilon^{*(\lambda)}p_H) (p_H+p_\rho)\nonumber\\
& & {} + \frac{iA_3(t)}{m_H + m_\rho} (\epsilon^{*(\lambda)}p_H)
(p_H-p_\rho) + \frac{2V(t)}{m_H + m_\rho}
\epsilon_\mu^{\phantom{\mu}\nu\rho\sigma}\epsilon_\nu^{*(\lambda)}
p_{H\rho} p_{\rho\sigma}.\makebox[0.8cm]{}
\end{eqnarray}
These are the relevant matrix-elements governing the hadronic part
of the
decays in question, decomposed in terms of the formfactors $f_\pm$,
$A_i$ and $V$, where $m_H$ and $m_\rho$ are the masses of the $H$
and the $\rho$ meson, respectively; $\lambda$ denotes the
polarization
state of the $\rho$. In the limit of vanishing lepton mass,
 the formfactors $f_-$ and $A_3$ do not contribute to the
decay rates and henceforth will not be considered. Expressed in terms
of the above formfactors, the spectra with respect to the electron
energy $E$ read:
\begin{eqnarray}
\lefteqn{\frac{d\Gamma(H\to\pi^+ e^-\bar \nu)}{dE} =
\frac{G_F^2}{16\pi^3m_H}\int\limits_0^{t_{max}}\!\!
dt \{ 2 E (m_H^2 - m_\pi^2 + t)-m_H(t + 4 E^2 )\}  f_+^2(t),}\\
\lefteqn{\frac{d\Gamma(H\to\rho^+ e^-\bar \nu)}{dE} = }\nonumber\\
& = & \frac{G_F^2}{128\pi^3m_H^2}\int\limits_0^{t_{max}}\!\! dt \, t
 \{(1 -
\cos \theta )^2 H_-^2 + (1 + \cos \theta)^2 H_+^2 + 2 (1 - \cos^2
\theta ) H_0^2\}\makebox[1.4cm]{}
\end{eqnarray}
with the helicity amplitudes
\begin{eqnarray}
H_\pm & = & (m_H + m_\rho) A_1(t) \mp
\frac{\lambda^{1/2}}{m_H+m_\rho} V(t),\\
H_0 & = & \frac{1}{2m_\rho\sqrt{t}}\left\{ (m_H^2 - m_\rho^2 - t)
(m_H
+ m_\rho) A_1(t) -
\frac{\lambda^{1/2}}{m_H + m_\rho}\,A_2(t)\right\}.
\end{eqnarray}
$t_{max}$, the maximum value of $t$, the invariant mass squared of
 the
lepton-pair, is given by
\begin{equation}
t_{max} = 2E \left( m_H - \frac{m_{\pi,\rho}^2}{m_H-2E} \right).
\end{equation}
$\theta$ is the angle between the $\rho$ and the charged lepton in
the
$(e^- \bar\nu)$ CM system and given by
\begin{equation}
\cos \theta = \frac{1}{\sqrt{\lambda}} \,( m_H^2 -m_\rho^2 + t - 4
m_H
E)
\end{equation}
where $\lambda = (m_H^2+m_\rho^2-t)^2-4 m_H^2 m_\rho^2$.

Returning to $(\ref{eq:corrpi})$ and
$(\ref{eq:corrga})$, it was the idea of Shifman, Vainshtein and
Zakharov \cite{SVZ} to account for non-perturbative corrections to
correlation functions by expressing them via an operator product
expansion (OPE) including terms that vanish in the perturbative
vacuum,
but acquire finite values in the QCD vacuum. These so-called
condensates characterize the long-distance behaviour of the
correlation function and we are led to write, e.g.
\begin{equation}\label{eq:OPE}
\Pi_+(p_H^2,p_L^2,t) = \sum_n \Pi_+^{(n)}(p_H^2,p_L^2,t)
\langle\, 0\,|\, {\cal O}_n\,| \, 0\, \rangle.
\end{equation}
The Wilson coefficients $\Pi_+^{(n)}$ can be calculated with the aid
of perturbation theory for negative values of $p_H^2-m_h^2$ and
$p_L^2-m_l^2$. The $\langle\, 0\,|\, {\cal O}_n\,| \, 0\, \rangle$
are
vacuum expectation values of gauge invariant operators, the
above-mentioned
condensates (cf.\ App.\ A). The first term in the series just covers
 usual
perturbation theory, the others are non-perturbative
corrections. In our analysis we will take into account the lowest
dimensional condensates up to dimension 6, where we improve existing
calculations \cite{alek} by the inclusion of the contribution of the
gluon condensate (App.~A). Equating
(\ref{eq:linkeseite}) and the OPE (\ref{eq:OPE}) yields expressions
for the formfactors determining (\ref{eq:corrpi}) and
(\ref{eq:corrga}) in terms of QCD parameters (like quark masses) and
condensates. Before, however, we can start to evaluate these sum
rules,
we have to specify how to treat $\rho^{\mbox{\scriptsize cont}}$ in
(\ref{eq:rhocont}). As for that, we employ the argument of
quark-hadron-duality (\cite{SVZ}, e.g.) and model
$\rho^{\mbox{\scriptsize cont}}$ by the contribution of usual
perturbation theory above some thresholds $s_H^0$ and $s_L^0$,
\begin{equation}
\rho^{\mbox{\scriptsize cont}} = \rho^{\mbox{\scriptsize pert}}\,
(1-\Theta (s_L^0-s_L)\,\Theta (s_H^0-s_H) ),
\end{equation}
which restricts the region of integration in the plane of $s_H$ and
$s_L$ to the gray area depicted in Fig.~\ref{fig:intgebiet}.
The calculation of $\rho^{\mbox{\scriptsize pert}}$ for $t>0$
involves
some delicate points connected with the possibility of the spectral
function to become singular. For the discussion of the additional
``non-Landau'' contributions caused by these singularities we refer
to
\cite{BBD}. In the numerical analysis we will tacitly include those
contributions whenever necessary.

The dependence of the sum rule on the continuum model as well as
the error induced by truncating the OPE series can be diminished by
the application of a Borel transformation. For an arbitrary function
of Euclidean momentum, $f(P^2)$ with $P^2=-p^2$, that
transformation is defined by
\begin{equation}\label{eq:defborel}
\hat{f} := \widehat B_{P^2}(M^2)\, f = \lim_{\begin{array}{c}
\scriptstyle
                  P^2\to\infty,N\to\infty\\[-1mm]
\scriptstyle      P^2/N = M^2\mbox{\scriptsize\ fixed}
                  \end{array}}
           \frac{1}{N!} (-P^2)^{N+1} \frac{d^{N+1}}{(dP^2)^{N+1}}\,
f,
\end{equation}
where $M^2$ is a new variable, called Borel parameter. For a typical
term appearing in the OPE, the transformation yields
\begin{equation}
\widehat B_{P^2}(M^2) \,\frac{1}{(p^2-m^2)^n} =
\frac{1}{(n-1)!}\, (-1)^n\, \frac{1}{(M^2)^n}
e^{-m^2/M^2}.
\end{equation}
Since condensates with high dimension get multiplied
by high powers of $(p^2-m^2)$ in the denominator, their contributions
get suppressed by factorials. In addition, the contribution of
higher resonances and the continuum,
$\rho^{\mbox{\scriptsize cont}}$,
gets exponentially suppressed relatively to the contribution of the
ground state, which is just the desired effect.

We are now in a position to write down sum rules for the relevant
formfactors. From (\ref{eq:corrpi}) and (\ref{eq:corrga}) we find:
\begin{eqnarray}
f_+^{H\to L}(t) & = & \frac{m_h}{f_Hf_\pi m_H^2}\, \exp \left\{
\frac{m_H^2}{M_h^2} + \frac{m_\pi^2}{M_u^2} \right\} M_h^2 M_u^2
\hat{\Pi}_+,\label{eq:srfplus}\\
A_1^{H\to L}(t) & = & \frac{m_h}{f_Hf_\rho (m_H+m_\rho) m_H^2 m_L}\,
\exp \left\{ \frac{m_H^2}{M_h^2} + \frac{m_\rho^2}{M_u^2} \right\}
M_h^2 M_u^2 \hat{\Gamma}_0,\label{eq:sra1}\\
A_2^{H\to L}(t) & = & \frac{m_h(m_H+m_\rho)}{f_Hf_\rho m_H^2m_L}\,
\exp \left\{ \frac{m_H^2}{M_h^2} + \frac{m_\rho^2}{M_u^2} \right\}
M_h^2 M_u^2 \hat{\Gamma}_+,\label{eq:sra2}\\
V(t)^{H\to L} & = & \frac{m_h(m_H+m_\rho)}{2f_Hf_\rho m_H^2m_L}\,
\exp \left\{ \frac{m_H^2}{M_h^2} + \frac{m_\rho^2}{M_u^2} \right\}
M_h^2 M_u^2 \hat{\Gamma}_V.\label{eq:srv}
\end{eqnarray}
Parts of the explicit formul\ae\ for the correlation functions can be
found in \cite{BBD}. For the present analysis, we in addition have
calculated the contributions of the gluon condensate and the
contribution of the four-quark condensate to $f_+$; the formul\ae\
can be found in the appendices. Note that we have expressed the
vacuum-to-meson transition amplitudes in terms of the corresponding
leptonic decay constants as
\begin{eqnarray}
\langle \, 0 | \, \bar d i \gamma_5 b \, | \, \bar B^0 \, \rangle &
= & f_B
\frac{m_B^2}{m_b},\\
\langle \, 0 \, | \, \bar d \gamma_\nu u \, | \, \rho^+, \lambda\,
\rangle & = & f_\rho m_\rho \epsilon^{(\lambda)}_\nu,\\
\langle \, 0 |\,  \bar d \gamma_\nu\gamma_5 u \, | \, \pi^+\,
\rangle
& = & i f_\pi p_{\pi\nu}.
\end{eqnarray}
The question of how to treat these quantities will be discussed in
the
next section.

\section{Evaluation of the sum rules}

In the numerical evaluation of the sum rules (\ref{eq:srfplus}) to
(\ref{eq:srv}) we use the following values of the condensates at a
renormalization scale of 1~GeV:
\begin{eqnarray}
\langle \bar q q \rangle (1\,\mbox{GeV}) & = &
(-0.24\,\mbox{GeV})^3,\nonumber\\
\langle \frac{\alpha_s}{\pi} G^2 \rangle & = & 0.012\,\mbox{GeV}^4\;
{},\nonumber\\
\langle\bar{q}\sigma gGq \rangle (1\,\mbox{GeV}) & = &
0.8\,\mbox{GeV}^2 \langle \bar q q \rangle (1\,\mbox{GeV}),
\nonumber\\
\pi\alpha_s \langle \bar q
      \gamma^\tau \lambda^A q \sum_{u,d,s}
      \bar q \gamma_\tau \lambda^A q \rangle & \approx & -
\frac{16}{9}\, \pi
\alpha_s \langle \bar q q \rangle^2,\nonumber\\
4\pi\alpha_s \langle \bar d \bar u u d \rangle & \approx & 4\pi
\alpha_s
\langle \bar q q \rangle^2.
\end{eqnarray}
We use leading-order anomalous dimensions of the quark and the mixed
condensate to evaluate them at a scale $\mu$ which is given by
the harmonic mean of the Borel parameters, $\mu^2 = \sqrt{M_h^2
M_u^2}$. For the four-quark condensates we assume vacuum saturation.
Since their contributions are tiny, we neglect the
scale-dependence. In general the sensitivity of the formfactors on
the
actual values of the condensates will be smaller than 10\% when
changing $(-\langle \bar q q \rangle)^{1/3}$ by 10~MeV and most
pronounced for the axial formfactors $A_2$. The  smallness of the
contributions
of the contributions of the four-quark condensates indicates that
higher order
power corrections are well under control.
Concerning quark masses, we put the masses of the
$u$ and the $d$ quark to zero, for the heavy quarks we use the
renormalization-group and -scheme invariant pole mass. Its connection
to the running mass in the $\overline{\mbox{MS}}$ scheme is given by
 (for
scales $\mu\ll m_{\overline{MS}}$)
\begin{equation}
m_{\mbox{\scriptsize pole}} = m_{\overline{MS}}(\mu) \left\{ 1 +
\frac{\alpha_s(\mu)}{\pi} \left( \frac{4}{3} + \ln
\frac{\mu^2}{m_{\overline{MS}}^2} \right) \right\}.
\end{equation}
The numerical values are (cf.\ \cite{narr}; a recent determination of
$m_b$ is given in \cite{dompfaffmb})
\begin{equation}
m_b = (4.6\mbox{--}4.8)\,\mbox{GeV},\qquad m_c =
(1.3\mbox{--}1.4)\,\mbox{GeV}.
\end{equation}
For the leptonic decay constants we use the experimental values
$f_\pi
= 0.133$~GeV and $f_\rho = 0.216$~GeV. For $f_B$ and $f_D$ we employ
two-point sum rules (\cite{alel}, e.g.), discarding radiative
corrections. We expect the accuracy of the
sum rules for the formfactors to be increased by that, since both in
the limit of infinitely heavy quarks and for the
matrix-element $\langle B | V_\mu | B \rangle$, where charge
conservation fixes the formfactor at zero recoil, QCD sum rules yield
the correct normalization independent of the values of quark masses,
continuum thresholds and Borel parameter \cite{ball,rady,neubert},
provided
the continuum thresholds in the two-point and the three-point sum
rules are chosen equal and the Borel parameter in the three-point sum
rule takes twice the value of that of the two-point sum rule. We will
take these prescriptions over to the case where the outgoing meson is
light, and actually the sensitivity of the resulting sum rules on
$m_b$ or $m_c$ is greatly reduced as compared to the sum rule for
$f_B$ or $f_D$. In
addition, the effect of the unknown radiative corrections to the
three-point function should tend to cancel against the radiative
corrections to the two-point function. Concluding, we take
both the range of Borel parameters, the ``sum rule window'', and the
values of the continuum thresholds from the two-point sum rules,
i.e.\
we evaluate (\ref{eq:srfplus}) to (\ref{eq:srv}) in the range
$7\,\mbox{GeV}^2\leq M_b^2\leq 10\,\mbox{GeV}^2$ and
$2\,\mbox{GeV}^2\leq M_c^2\leq 4\,\mbox{GeV}^2$ and for continuum
thresholds
$s_D^0 = (6\mbox{--}7)\,\mbox{GeV}^2$, $s_B^0 =
(34\mbox{--}36)\,\mbox{GeV}^2$, $s_\pi^0 = (0.75-1.0)\,{\rm GeV}^2$,
$s_\rho^0 = (1.25-1.5)\,{\rm GeV}^2$.
In addition, we choose a fixed ratio of the Borel parameters,
\begin{equation}
\frac{M_b^2}{M_u^2} = 4,\quad \frac{M_c^2}{M_u^2} = 2.
\end{equation}
This procedure ensures that perturbative and non-perturbative
corrections in both the heavy and the light channel are equally
weighted. The sum rules are rather insensitive to the actual value of
that ratio, and changing for example $M_b^2/M_u^2$ from 3 to 5
results
in changes of the results of at most 10\%.

Our aim is to extract the $t$-dependence of the formfactors from the
sum rules, and thus we are restricted to a range of values where the
correlation function can be expected to be reliable in that variable.
That is, we have to stay approximately 1~GeV$^2$ below
the perturbative cut starting at $t=m_{b,c}^2$. Thus we can
trust the sum rules up to $t \approx 20\,\mbox{GeV}^2$ for the
$B$-decays
and $t \approx 0.9\,\mbox{GeV}^2$ for the $D$-decays. The maximum
values
allowed by kinematics are $t_{max}^{B\to\pi} = 26.4\,\mbox{GeV}^2$,
$t_{max}^{B\to\rho} = 20.3\,\mbox{GeV}^2$, $t_{max}^{D\to\pi} =
3.0\,\mbox{GeV}^2$ and $t_{max}^{D\to\rho} = 1.2\,\mbox{GeV}^2$, so
apart from $B\to\rho$ we cannot cover the full range of $t$. Albeit
the accessible range is sufficient to determine the shape of the
formfactor
and the total rate, it
is not for the calculation of the electron spectrum. We thus take the
attitude to extrapolate the formfactors to $t_{max}$ to get the
electron spectrum for large values of the electron energy $E$. This
procedure does not introduce too large an uncertainty for $D\to\rho$
since according to the above remarks at least 80\% of the integration
range in $t$ is covered, and for $D\to\pi$ and $B\to\pi$ we will find
a pole-type behaviour of the formfactors which facilitates the
applicability of the extrapolation.

In Fig.~\ref{fig:fplus0}(a) the formfactor $f_+^{D\to\pi}(t=0)$ is
shown as function of the Borel parameter $M_c^2$. The different
curves
correspond to different choices of the set of input parameters. To be
specific, we use $m_c = 1.3\,{\rm GeV}$, $s_D^0=6\,{\rm GeV}^2$,
$s_\pi^0 = 0.75\,{\rm GeV}^2$ (set C1), $m_c = 1.3\,{\rm GeV}$,
$s_D^0=6\,{\rm GeV}^2$, $s_\pi^0 = 1\,{\rm GeV}^2$ (set C2), $m_c =
1.4\,{\rm GeV}$, $s_D^0=7\,{\rm GeV}^2$, $s_\pi^0 = 0.75\,
{\rm GeV}^2$
(set C3), $m_c = 1.4\,{\rm GeV}$, $s_D^0=7\,{\rm GeV}^2$, $s_\pi^0 =
1\,{\rm GeV}^2$ (set C4). The value of $s_D^0$ is taken as the
best-fit continuum threshold for the sum rules for $f_D$, $s_\pi^0$
is
taken from \cite{SVZ}. In the ``sum rule window'' $2\,
\mbox{GeV}^2\leq
M_c^2 \leq
4\,\mbox{GeV}^2$ the $f_+^{D\to\pi}(0)$ is quite stable and the
dependence on
the values of both the mass of the c-quark and the continuum
thresholds
as respresented by the spread of curves is well under control,
yielding $f_+(0) = 0.5\pm 0.1$ where the error is an educated guess
based on both the dependence of the sum rule on the input parameters
and the intrinsic uncertainty of the whole method. Perturbation
theory
and quark condensate give the dominant contribution, the other
condensates contributing at the level of $\sim$10\%. Thus the series
of power corrections is well under control.

$f_+^{B\to \pi}(0)$ as function of the Borel parameter $M_b^2$ is
shown
in Fig.~\ref{fig:fplus0}(b). Here we use the parameter sets B1
($m_b =
4.6\,{\rm GeV}$, $s_B^0=36\,{\rm GeV}^2$, $s_\pi^0 = 0.75\,{\rm
GeV}^2$), B2 ($m_b = 4.8\,{\rm GeV}$, $s_B^0=36\,{\rm GeV}^2$,
$s_\pi^0 = 1\,{\rm GeV}^2$), B3 ($m_b = 4.8\,{\rm GeV}$, $s_B^0=34\,
{\rm GeV}^2$, $s_\pi^0 = 0.75\,{\rm GeV}^2$), B4 ($m_b = 4.8\,{\rm
GeV}$, $s_B^0=34\,{\rm GeV}^2$, $s_\pi^0 =  1\,{\rm GeV}^2$). Again
the
value is remarkably stable against variation in the quark mass, in
the
continuum thresholds and the Borel parameter. Form
Fig.~\ref{fig:fplus0}(b) we find $f_+^{B\to \pi}(0) = 0.26\pm 0.02$.
This value is higher than obtained in \cite{BPi} which is
due to the contribution of the gluon condensate not included there.

In Fig.~\ref{fig:fplust}(a) we show $f_+^{D\to\pi}(t)$ as function of
$t$, normalized to its value at $t=0$, which representation
emphasizes
the differences in shape. We have chosen $M_c^2 = 3\,{\rm GeV}^2$ and
give curves for all parameter sets. We find a rise in $t$ which is
very well compatible with a pole-type behaviour as suggested by the
pole dominance hypothesis (\ref{eq:poldominanz}). From a pole fit
we get $m_{pol} = (1.95\pm 0.10)\,{\rm GeV}$ (including all sets and
Borel parameters within the window). If pole-dominance were exactly
valid, the pole-mass would be $m_{D^*} = 2.01\,{\rm GeV}$, so QCD
sum rules
confirm pole-dominance for $D\to\pi$.

Pole-dominance is likewise valid for the $B\to\pi$ transition whose
normalized
formfactor is depicted in Fig.~\ref{fig:fplust}(b) as function of $t$
and for all parameter sets at $M_b^2 = 8\,{\rm GeV}^2$. Pole fits
yield pole-masses of $\sim$5.1~GeV for the sets B1 and B2 and
$\sim$5.2~GeV for B3 and B4. From that we are forced to exclude the
lower value of the b-quark mass, $m_b = 4.6\,{\rm GeV}$, from our
analysis (since the formfactor would become singular at $t_{max}$)
and
stick to $m_b = 4.8\,{\rm GeV}$. The ``physical'' pole is at $m_{B^*}
= 5.33\,{\rm GeV}$ what nicely agrees with the fit value
$(5.25\pm0.10)\,$GeV from sets B3 and B4.

In Fig.~\ref{fig:specpi} we show the electron spectra $d\Gamma/dE$ as
functions of the electron energy $E$ which can be obtained from the
formfactors $f_+^{D\to\pi}$ and $f_+^{B\to\pi}$, extrapolated up to
$t_{max}$ according to pole-dominance.

Let us now turn turn to the decays $H\to\rho e \nu$. In
Fig.~\ref{fig:FF0D} we show the formfactors of $D\to\rho$ at $t=0$ as
functions of the Borel parameter for the parameter sets C5 ($m_c =
1.3\,{\rm GeV}$, $s_D^0=6\,{\rm GeV}^2$, $s_\rho^0 = 1.25\,{\rm
GeV}^2$), C6 ($m_c = 1.3\,{\rm GeV}$, $s_D^0=6\,{\rm GeV}^2$,
$s_\rho^0 = 1.5\,{\rm GeV}^2$), C7 ($m_c = 1.4\,{\rm GeV}$,
$s_D^0=7\,
{\rm GeV}^2$, $s_\rho^0 = 1.25\,{\rm GeV}^2$), C8 ($m_c = 1.4\,{\rm
GeV}$, $s_D^0=7\,{\rm GeV}^2$, $s_\rho^0 = 1.5\,{\rm GeV}^2$). All
formfactors are quite stable and we find $A_1^{D\to\rho}(0) = 0.5\pm
0.2$, $A_2^{D\to\rho}(0) = 0.4\pm 0.1$ and $V^{D\to\rho}(0) = 1.0\pm
0.2$ where as in the previous cases the error is intended to include
likewise systematic uncertainties.

For $B\to\rho$ we find from Fig.~\ref{fig:FF0B} $A_1^{B\to\rho}(0) =
0.5\pm 0.1$, $A_2^{B\to\rho}(0) = 0.4\pm 0.2$, $V^{B\to\rho}(0) =
0.6\pm 0.2$
with the parameter sets B5 ($m_b =
4.6\,{\rm GeV}$, $s_B^0=36\,{\rm GeV}^2$, $s_\rho^0 = 1.25\,{\rm
GeV}^2$), B6 ($m_b = 4.6\,{\rm GeV}$, $s_B^0=36\,{\rm GeV}^2$,
$s_\rho^0 = 1.5\,{\rm GeV}^2$), B7 ($m_b = 4.8\,{\rm GeV}$,
$s_B^0=34\,
{\rm GeV}^2$, $s_\rho^0 = 1.25\,{\rm GeV}^2$), B8 ($m_b = 4.8\,{\rm
GeV}$, $s_B^0=34\,{\rm GeV}^2$, $s_\rho^0 = 1.5\,{\rm GeV}^2$). All
these formfactors depend only slightly on quark masses and
continuum thresholds and are stable in the Borel parameter, except
for
$A_2^{B\to\rho}(0)$. Here we observe for B7 and B8 a rather strong
dependence on $M_b^2$ the reason being the extremely small
contribution of
perturbation theory which is of about 10\% only.

In Fig.~\ref{fig:FFtD} the normalized formfactors of the $D\to\rho$
transition
are plotted as functions of $t$ for $M_c^2 = 3\,{\rm GeV}^2$ and all
 sets of
parameters. We find a {\em decrease} of $A_1^{D\to\rho}(t)$ in $t$
which is nearly
independent of the parameter set used. This behaviour is in clear
contradiction
with pole-dominance which predicts an increase determined by the
pole-mass
$m_{D^{1+}} = 2.42\,{\rm GeV}$ corresponding to
$A_1^{D\to\rho}(t)/A_1^{D\to\rho}(0) = 1.21$. A similar behaviour
is encountered for $A_2^{D\to\rho}$ where we find a
decrease of about 10\% at $t=1\,{\rm GeV}^2$ depending on the
parameter set used.
For the vector formfactor we have an increase in $t$ with a best-fit
pole-mass of $m_{pol} = (2.5\pm 0.2)\,{\rm GeV}$ which is a little
bit larger
than predicted by pole-dominance.

For the normalized formfactor $A_1^{B\to\rho}(t)/A_1^{B\to\rho}(0)$,
 shown in
Fig.~\ref{fig:FFtB}(a) at $M_b^2 = 8\,{\rm GeV}^2$ for all parameter
 sets, we
find a rather unexpected shape with a minimum at $t\approx 15\,
{\rm GeV}^2$.
Formally, this minimum is due to the interplay between decreasing
contributions
of perturbation theory and quark condensate and an increasing one of
 the gluon
condensate which becomes effective at large $t$. For
$A_2^{B\to\rho}(t)/
A_2^{B\to\rho}(0)$ (Fig.~\ref{fig:FFtB}(b)) we find a moderate
increase in $t$
which at large $t$ is again compensated by a negative contribution
of the gluon
condensate. For $V^{B\to\rho}(t)/V^{B\to\rho}(0)$
Fig.~\ref{fig:FFtB}(c) shows
the usual increase in $t$ corresponding to a pole mass of
$(6.6\pm 0.6)\,{\rm GeV}$,
about $1\,{\rm GeV}$ larger than predicted by pole-dominance.

Finally, in Fig.~\ref{fig:specDrho} we show the electron-spectrum
$d\Gamma/dE$ of the decay $D\to\rho e \nu$ as function of the
electron energy $E$ for set C5 and $M_c^2 = 3\,{\rm GeV}^2$ where
the formfactors are extrapolated in the range
$t\geq 1\,{\rm GeV}^2$. Fig.~\ref{fig:specBrho}(a) shows the
electron spectrum
$d\Gamma/dE$ of $B\to\rho e \nu$ as function of $E$ for set B7 and
$M_b^2 = 8\,
{\rm GeV}^2$ as quite sharp and concentrated around large electron
energies.

\section{Results and Discussion}

In the previous section we have given a careful analysis of the
semileptonic heavy-light decays $D\to\pi e \nu$, $B\to\pi e \nu$,
$D\to\rho e \nu$ and $B\to\rho e \nu$. We have put some stress on the
calculation of the $t$-dependence of the formfactors which for the
vector formfactors in general can well be described by a
pole-dominance formula, whereas the axial formfactors tend to
decrease
in $t$ and even develop extrema. The numerical results of our
calculation as well as other models are collected in the tables. The
formfactors at $t=0$ can be found in Tables~I and II, the rates
in Tables~III and IV. The rates were calculated either using
pole-dominance (as indicated in the tablenotes) or some other model
for the $t$-dependence. In addition to the total rates we give for
$H\to
\rho e \nu$ the ratios $\Gamma_L/\Gamma_T$ and $\Gamma_+/\Gamma_-$
where the index denotes the polarization state of the $\rho$
(longitudinal, transversal, positive, and negative helicity,
respectively). The corresponding electron spectra are shown in
Figs.~\ref{fig:specpi} ($D,B\to \pi e \nu$), \ref{fig:specDrho}
($D\to\rho e \nu$) and \ref{fig:specBrho}(a) ($B\to\rho e \nu$).

The only decay, where a comparison to experiment is possible so far,
is $D\to\pi e \nu$. In addition there exist several model
calculations in literature, using QCD sum rules
\cite{unpub,dompfaffD}, quark models \cite{BWS,KS,GISW} and some
lattice calculations \cite{LMS,BKS}. One calculation relying on the
heavy quark effective theory \cite{gatto} takes the experimental
result \cite{pardat} as input to their values of the formfactors of
$B\to\pi,\,\rho e \nu$. The theoretical predicitions of
$\Gamma(D\to\pi e\nu)$ differ by a factor of two, and assuming
$|V_{cd}| = 0.22$, which can be inferred from the unitarity of the
CKM-matrix with high accuracy, we find that the central value of our
rate is by two
standard deviations smaller than the experimental value. This
discrepancy is not strong enough to be conclusive and might be
due to the neglection of radiative corrections to our sum sules.
Still
further experimental effort in improving statistics is to be
desired to
clarify this point.

For $D\to\rho e \nu$ experiment only has set an upper bound for the
total rate so far
\cite{slobo}. Our value is by a factor five smaller than these. For
the ratio $\Gamma(D\to\rho)/\Gamma(D\to\pi)$ we get 0.3 which again
is
smaller than the predictions in other models which yield a maximum
values of 1.8 \cite{GISW}. We remind that the corresponding ratio for
the Cabibbo favoured decays, $\Gamma(D\to K^*)/\Gamma(D\to K)$, is
approximately 0.5 \cite{stone} and that we do not expect flavour
SU(3)
to be broken by a factor of two or more. The formfactors at $t=0$
roughly agree in all models except for the lattice calculation
\cite{BKS} which predicts vanishing $A_2^{D\to\rho}(0)$ and a small
value of $V^{D\to\rho}(0)$ yielding a large value of
$\Gamma_L/\Gamma_T$.

For the $b\to u$ decays we do not dare to quote any experimental
upper
bound for the total rates due to the uncertainty in $|V_{ub}|$ (but
cf.\ \cite{CLEO}). We remark that the total rates for $B\to\pi e \nu$
summarized in Table~IV and obtained by QCD sum rules
\cite{slobo,narrdecay,dompfaffB}, the quark models \cite{BWS,KS,GISW}
and the HQET calculation \cite{gatto} differ by a factor of 26. For
$B\to\rho e \nu$ this value shrinks to 4. That spread in predictions
clearly shows the neccesity for an accurate investigation of the
$t$-dependence of the formfactors that we have concentrated on in
this
paper. With the $t$-dependence obtained by
QCD sum rules we obtain $\Gamma(\bar{B^0}\to \rho^+ e \bar\nu ) =
(1.2\pm
0.4)\,|V_{ub}|^2\cdot 10^{13}{\rm s}^{-1}$ where the $\rho$ has
mainly negative helicity. Furthermore, we find
$\Gamma(B\to\rho)/\Gamma(B\to\pi) = 2.4$ which is smaller than all
other model predictions ranging from 3.1 to 11 except for
\cite{gatto}
which predicts 0.6. In Fig.~\ref{fig:specBrho}(b) we give the
electron
spectrum $1/\Gamma\,d\Gamma/dE$ for $B\to\rho e \nu$ as obtained in
this paper (same parameters as in Fig.~\ref{fig:specBrho}(a)), in the
BWS model \cite{BWS} (using pole-dominance) and the non-relativistic
 GISW model \cite{GISW}.
The chosen normalization emphasizes the difference in shape rather
than in the absolute normalization. Although the BWS spectrum is
softer in the endpoint region above the threshold for
charm-production, the only region where $b\to u$ transitions can be
observed, our spectrum and that of
GISW are nearly indistinguishable for $E\leq 2.4\,{\rm GeV}$. If,
however, a detection of the polarization of the $\rho$ was feasible,
one could test the considerable deviations of the corresponding
spectra in the different models. We predict the $\rho$ to have
predominantly negative helicity (as indicated by the very small
values
of $\Gamma_L/\Gamma_T$ and $\Gamma_+/\Gamma_-$) whereas in other
models the ratio $\Gamma_L/\Gamma_T$ is closer to one.

In Fig.~\ref{fig:inklusiv} we give a comparison of the inclusive
$b\to
u$ semileptonic spectrum calculated by means of QCD sum rules in
\cite{inklupreprint} to the exclusive decay spectra $B\to\pi e \nu$
and $B\to\rho e \nu$ calculated with the same parameters ($m_b =
4.8\,$GeV, $s_B^0 = 34\,{\rm GeV}^2$, $M_b^2 = 8\,{\rm GeV}^2$).
Fig.~\ref{fig:inklusiv}(a) shows the spectrum in the restframe of the
decaying B-meson, Fig.~\ref{fig:inklusiv}(b) in the laboratory system
of an $e^+ e^-$ collider operating on the $\Upsilon(4S)$ resonance.
{}From both we find that at high electron energies $B\to\rho e \nu$
constitutes nearly
the whole differential inclusive rate, so it is worthwile to
concentrate on measurements of the exclusive channels, where
theoretical predictions are still not at their best precision, but
are
much more accurate than calculations of the inclusive spectrum (cf.\
\cite{inklupreprint}).

\acknowledgments

It is a pleasure to thank V.M.\ Braun and H.G.\ Dosch for useful
discussions.

\appendix{The Wilson coefficient of the gluon
condensate}

In the following we present a technique for calculating
Borel transformed Wilson coefficients directly from the
loop-integrals. This method does not allow for the subtraction of
continuum contributions, which, however, does no harm in our case as
the total contribution of the gluon condensate to the three-point sum
rule is small by itself ($\alt {\cal O}(10\%)$), and so is its
continuum portion. Besides, one would expect typical continuum
contributions to show up as incomplete Gamma functions in the Wilson
coefficiens, which, however, are absent in our formul\ae\ (i.e.\ in
the sum of all diagrams, but are encountered in each diagram
separately). Thus one is led to conclude that those contributions are
actually absent in the processes under consideration.

We calculate the diagrams shown in Fig.~\ref{fig:diagrams} that
contribute to the Wilson-coefficient of the gluon condensate
$\langle
\alpha_s G^2/\pi\rangle$ in the fixed point gauge
\begin{equation}
x^\mu \, A^{a}_\mu(x) = 0
\end{equation}
with the gluon field $A^{a}_\mu$, $a\in \{ 1,2,\dots,8\}$. For
massless light quarks and with the coordinates chosen as indicated in
the first diagram, diagrams I and II evaluate to zero. Note, that for
massless light quarks there is no mixing of the gluon with the quark
condensate.

In the evaluation of the remaining diagrams we encounter integrals
of type
(since the Borel transform removes UV divergencies, there is no need
for dimensional regularization of these divergencies and we thus stay
with four-dimensional integrals)
\begin{equation}
I_{\mu_1\mu_2\cdots\mu_n}(a,b,c) = \int\!\!\frac{d^4\! k}{(2\pi)^4}\,
\frac{k_{\mu_1} k_{\mu_2} \cdots k_{\mu_n}}{[k^2]^a [(k+p_\rho)^2]^b
[(k+p_B)^2-m_b^2]^c}.
\end{equation}
Although the sum of all diagrams is IR convergent, IR divergent terms
occur at each step of the calculation and need proper regularization.
We take the attitude to let the mass of the u-quark finite in the
denominator of its propagator, $m_u^2>0$ (but let $m_u=0$ in the
traces) and regularize the singularities in the q-quark line (for
$a=2$) by shifting the power of the q-quark propagator to
$2-\epsilon$
(which from a technical point of view is simpler than introducing
dimensional regularization). This procedure has got the
advantage that all integrals with $b,\,c>1$ can be obtained from the
case $b=c=1$ by taking derivatives with respect to the quark masses:
\begin{equation}
I_{\mu_1\mu_2\cdots\mu_n}(a,b,c) = \frac{1}{\Gamma(b)\Gamma(c)}\,
\frac{d^{b-1}}{d(m_u^2)^{b-1}}\,\frac{d^{c-1}}{d(m_b^2)^{c-1}}\,
I_{\mu_1\mu_2\cdots\mu_n}(a,1,1).
\end{equation}
Continuing to Euclidean space--time and employing the Schwinger
representation for propagators,
\begin{equation}
\frac{1}{[P^2+m^2]^a} = \frac{1}{\Gamma(a)}\int\limits_0^\infty
\!\! d\alpha\, \alpha^{a-1}\, e^{-\alpha (P^2+m^2)},
\end{equation}
we find for the scalar integral $n=0$ with $a=b=c=1$ (with capital
letters denoting Euclidean momenta):
\begin{equation}
I(1,1,1) = -i\!\! \int\limits_0^\infty \!\!
d\alpha\,d\beta\,d\gamma\!\int\!\!\frac{d^4\!
\widetilde{K}}{(2\pi)^4}\,
\exp ( -\Sigma \widetilde{K}^2 -\frac{\alpha\beta}{\Sigma}P_\rho^2
-\frac{\alpha\gamma}{\Sigma} P_B^2 -\frac{\beta\gamma}{\Sigma} T
-\gamma m_b^2)
\end{equation}
where
\begin{mathletters}
\begin{eqnarray}
\widetilde{K} & = & K + \frac{1}{\Sigma} (\beta P_\rho +
\gamma P_B),\\
\Sigma & = & \alpha + \beta + \gamma,\\
T & = & -t.
\end{eqnarray}
\end{mathletters}
The above representation proves very convenient for applying the
Borel
transformation with
\begin{equation}
\widehat{B}_{P^2}(M^2)\, e^{-\alpha P^2} = \delta (1-\alpha M^2).
\end{equation}
{}From that, we get
\begin{eqnarray}
\hat{I}(1,1,1) & := & \widehat{B}_{P_\rho^2}(M_u^2)\,
\widehat{B}_{P_B^2}(M_b^2) I(1,1,1)\nonumber\\
 & = & \frac{i}{16\pi^2}\,\frac{1}{M_b^2+M_u^2}\,\,
e^{-t/(M_b^2+M_u^2)}\, {\rm Ei}(-z)
\end{eqnarray}
where the exponential integral function is given by
\begin{equation}
{\rm Ei}(x) = -\int\limits_{-x}^{\infty}\!\! dt\,\frac{e^{-t}}{t}
\end{equation}
and
\begin{equation}
z = \frac{m_u^2}{M_u^2} + \frac{m_b^2}{M_b^2} -
\frac{t}{M_b^2+M_u^2}.
\end{equation}
Actually IR divergent diagrams only occur for the scalar case where
we find
\begin{equation}
\hat{I}(2-\epsilon,1,1) = \frac{i}{16\pi^2}\,
\frac{e^{-m_u^2/M_u^2-m_b^2/M_b^2}}{M_b^2 M_u^2}
\left\{ \frac{1}{\epsilon} + 1 - 2
\gamma_E - \ln \left(-\frac{\mu^2}{M_u^2}-
\frac{\mu^2}{M_b^2}\right) -
\ln z\right\}.
\end{equation}
Here $\mu$ is some arbitrary scale introduced to render the canonical
dimension of the integral, which, however, cancels in the complete
expressions for Wilson-coefficients, as it should.

For larger values of $n$, we get
\begin{eqnarray}
\hat{I}_{\mu_1}(1,1,1) & = & \frac{-i}{16\pi^2}
\left( \frac{M_b^2 M_u^2}{M_b^2+M_u^2} \right)^2
\frac{1}{M_u^2 M_b^2} \left(\frac{p_{\rho\mu_1}}{M_u^2} +
\frac{p_{B\mu_1}}{M_b^2}\right)\nonumber\\
& & \left( e^{-m_b^2/M_b^2-m_u^2/M_u^2} + e^{-t/(M_b^2+M_u^2)}
(1+ z\, {\rm Ei}(-z)) \right),\\
\hat{I}_{\mu_1}(2,1,1) & = & \frac{i}{16\pi^2} \frac{1}{M_b^2+M_u^2}
\left(\frac{p_{\rho\mu_1}}{M_u^2} + \frac{p_{B\mu_1}}{M_b^2}\right)
e^{-t/(M_b^2+M_u^2)} \, {\rm Ei}(-z),\\
\hat{I}_{\mu_1\mu_2}(a\leq 2,1,1) & = & \frac{i}{16\pi^2}
\frac{(-1)^a}{\Gamma(a)} \left(\frac{M_b^2 M_u^2}{M_b^2+M_u^2}
\right)^{4-a}
\frac{e^{-m_b^2/M_b^2-m_u^2/M_u^2}}{M_u^2 M_b^2} \,\Gamma(4-a)
\nonumber\\
& & \left\{ -\frac{1}{2(3-a)}\left( \frac{1}{M_b^2} +
\frac{1}{M_u^2}\right)
g_{\mu_1\mu_2} \,U\left(3-a,0;z\right)\right.\nonumber\\
& & {}+ \left( \frac{p_{\rho\mu_1}p_{\rho\mu_2}}{M_u^4} +
\frac{p_{\rho\mu_1}p_{B\mu_2} +
p_{B\mu_1}p_{\rho\mu_2}}{M_b^2M_u^2} +
\frac{p_{B\mu_1}p_{B\mu_2}}{M_b^4}\right)\nonumber\\
& & {}\left. \vphantom{\frac{p_{\rho\mu_1}p_{\rho\mu_2}}{M_u^4}}
U\left(4-a,1;z\right)\right\},
\end{eqnarray}
where we have continued back to Minkowski--space. $U$ is the
confluent
hypergeometric function defined as
\begin{equation}
U(i,j;x) = \frac{1}{\Gamma(i)}\int\limits_0^\infty\!\!
dt\,(1+t)^{j-i-1}\, t^{i-1}\, e^{-xt}.
\end{equation}
In addition, we use
\begin{eqnarray}
\lefteqn{\widehat{B}_{P_\rho^2}(M_u^2)\,\widehat{B}_{P_B^2}(M_b^2)\,
[p_\rho^2]^{m_1} [p_B^2]^{m_2}
I_{\mu_1\mu_2\cdots\mu_n}(a,b,c)\ =}\nonumber\\
& = &  [M_u^2]^{m_1} [M_b^2]^{m_2} \,\frac{d^{m_1}}{d
(M_u^2)^{m_1}}\,\frac{d^{m_2}}{d (M_b^2)^{m_2}}\, [M_u^2]^{m_1}
[M_b^2]^{m_2} \hat{I}_{\mu_1\mu_2\cdots\mu_n}(a,b,c).
\end{eqnarray}
We now decompose the Lorentz-invariants $\hat\Lambda$ occuring in
the
Borel transformed correlation functions (\ref{eq:corrpi}) and
(\ref{eq:corrga}) as
\begin{equation}\label{eq:decomposition}
\hat{\Lambda} = \sum_n \hat{\Lambda}^{(n)} \langle{\cal O}_n\rangle
\end{equation}
where $\langle{\cal O}_1\rangle = \langle\openone\rangle = 1$,
$\langle{\cal O}_3\rangle = \langle\bar{q}q\rangle$,
$\langle{\cal O}_4\rangle = \langle\alpha_s G^2/\pi\rangle$,
$\langle{\cal O}_5\rangle = \langle\bar{q}\sigma gGq \rangle$,
$\langle{\cal O}_{6_1}\rangle = \pi\alpha_s \langle \bar q
	\gamma^\tau \lambda^A q \sum_{u,d,s}
	\bar q \gamma_\tau \lambda^A q \rangle$
and $\langle{\cal O}_{6_2}\rangle = 4\pi\alpha_s \langle \bar d
\bar u u d
\rangle$ are the condensates taken into account. The
formul\ae\ for $n\in \{1,3,5,6\}$ can be found in \cite{BBD} (with
the
same notations), where vacuum saturation for the condensates
with dimension 6 is assumed; the formul\ae\ for $n=4$ are new and
read
for the relevant invariants:
\begin{eqnarray}
\hat{\Pi}^{(4)}_+ & = &
\frac{m_be^{-m_b^2/M_b^2}}{M_b^2 M_u^2} \left[
\frac{1}{96M_b^2} + \frac{1}{48M_u^2}
- \frac{m_b^2 M_u^2\{ M_b^2 (M_b^2+3M_u^2)-m_b^2 (M_b^2+M_u^2)\}}{24
	M_b^6 (M_b^2+M_u^2)^2 z^3} \right.\nonumber\\
& & {}- \frac{8M_b^4 M_u^4 + 4 m_b^2 M_b^2
(M_b^4+4M_b^2M_u^2+2M_u^4) - m_b^4
	M_u^2 (2M_b^2 + M_u^2)}{96M_b^6(M_b^2 + M_u^2)^2z^2}
\nonumber\\
& & {}\left.+\frac{4M_b^2 (M_b^2+M_u^2) -
m_b^2(2M_b^2+M_u^2)}{48M_b^4
	(M_b^2+M_u^2)z}
+\frac{m_b^4 M_u^4}{16M_b^6(M_b^2+M_u^2)^2z^4}
\right],\\
\hat{\Gamma}^{(4)}_0 & = &
\frac{m_b e^{-m_b^2/M_b^2}}{M_b^2M_u^2} \left[
-\frac{m_b^6 M_u^6}{4M_b^8(M_u^2+M_b^2)^2z^5}
+ \frac{m_b^4 M_u^4 \{ M_b^2 ( 3 M_b^2 + 10 M_u^2) - 3 m_b^2 M_u^2
	\}}{16M_b^8(M_u^2+M_b^2)^2 z^4}\right.
\nonumber\\
& & {}+\frac{m_b^2 M_u^4 \{ m_b^2 M_b^2 (23M_u^2+7M_b^2) -2M_b^4
     (13M_u^2+5M_b^2)-3m_b^4M_u^2\}}{48M_b^8(M_u^2+M_b^2)^2z^3}
\nonumber\\
& & {}+\frac{M_u^2\{8M_b^6M_u^2(M_b^2+2M_u^2)+4m_b^2M_b^4
(M_b^4-4M_b^2M_u^2
	-11M_u^4)-m_b^6M_u^4\}}{96M_b^8(M_u^2+M_b^2)^2z^2}\nonumber\\
& & {}+\frac{m_b^4M_u^4(7M_b^2+16M_u^2)}{96M_b^6(M_u^2+M_b^2)^2z^2}
+\frac{M_u^2\{4M_b^2(2M_u^2-M_b^2)+m_b^2(2M_b^2-9M_u^2)
	\}}{48M_b^4(M_u^2+M_b^2)z}\nonumber\\
& & \left. {} + \frac{m_b^4M_u^4}{32M_b^6(M_u^2+M_b^2)z}
+\frac{M_b^2(8M_u^2-5M_b^2)-3m_b^2M_u^2}{96M_b^4}
+\frac{(M_b^2+M_u^2)z}{96M_b^2}
\right],\\
\hat{\Gamma}^{(4)}_+ & = &
\frac{m_b e^{-m_b^2/M_b^2}}{M_b^2M_u^2} \left[
\frac{1}{96M_b^2}
+ \frac{m_b^2M_u^2(M_b^2+3M_u^2)}{24M_b^4(M_b^2+M_u^2)^2z^3}
+ \frac{M_u^2(4M_b^2-m_b^2)}{48M_b^4(M_b^2+M_u^2)z}\right.\nonumber\\
& & \left. {}+
\frac{M_u^4(m_b^4-4m_b^2M_b^2-8M_b^4)}{96M_b^6(M_b^2+M_u^2)^2z^2}
-\frac{m_b^4M_u^4}{16M_b^6(M_b^2+M_u^2)^2z^4}
\right],\\
\hat{\Gamma}^{(4)}_V & = &
\frac{m_b e^{-m_b^2/M_b^2}}{M_b^2M_u^2} \left[
\frac{m_b^4M_u^4}{8M_b^6(M_b^2+M_u^2)^2z^4}
+ \frac{m_b^2M_u^2\{m_b^2 M_u^2 -
      M_b^2(M_b^2+3M_u^2)\}}{12M_b^6(M_b^2+M_u^2)^2z^3}\right.
\nonumber\\
& & {}+\frac{M_u^2\{8M_b^4M_u^2-4m_b^2M_b^2(M_b^2+2M_u^2)+
m_b^4M_u^2\}}{48M_b^6
      (M_b^2+M_u^2)^2z^2}
+ \frac{1}{48M_b^2}\nonumber\\
& & \left. {}
+ \frac{2M_b^2(2M_u^2-M_b^2)-m_b^2M_u^2}{24M_b^4(M_b^2+M_u^2)z}
\right].
\end{eqnarray}
Note that we have checked our method for calculating
Borel-transformed
Wilson-coefficients for the matrix element
$\langle\, B\, | \, V_\mu\, | \, B \,\rangle$
at $t=0$ where the result is uniquely determined by charge
conservation.

\appendix{The Wilson coefficient of the\\ four-quark condensate}

In addition to the formul\ae\ given in \cite{BBD}, we also have
calculated the contributions of the four-quark condensate to the
decays $B,\, D\to \pi e \nu$ which read (in the notation of
(\ref{eq:decomposition}):
\begin{eqnarray}
\hat{\Pi}_+^{(6_1)} & = & \frac{m_be^{-m_b^2/M_b^2}}{M_b^2 M_u^2}
\left( \frac{1}{9M_b^2M_u^2} - \frac{m_b^2}{72M_b^6}
- \frac{1}{18M_b^4} + \frac{t}{36M_b^4M_u^2} -
\frac{m_b^2-t}{18M_b^2M_u^4} \right),\nonumber\\
\hat{\Pi}_+^{(6_2)} & = & \frac{m_be^{-m_b^2/M_b^2}}{M_b^2 M_u^2}
\left( \frac{1}{9M_u^4} + \frac{2}{9M_u^2(m_b^2-t)} \right).
\end{eqnarray}

\begin{table}
\caption{The formfactors of the $c\to d$ transitions at $t=0$ in
different models.}
\begin{tabular}{lllll}
Reference & $f_+^{D\to \pi}$ &
$A_1^{D\to \rho}$ & $A_2^{D\to \rho}$ &
$V^{D\to \rho}$ \\ \tableline
This paper & 0.5$\pm$0.1 & 0.5$\pm$0.2 & 0.4$\pm$0.1 & 1.0$\pm$0.2\\
\cite{unpub}$^a$ & 0.7$\pm$0.2 & -- & -- & --\\
\cite{dompfaffD}$^a$ & 0.75$\pm$0.05 & -- & -- & --\\
\cite{BWS}$^b$ & 0.69 & 0.78 & 0.92 & 1.23 \\
\cite{GISW}$^b$ & 0.51 & 0.59 & 0.23 & 1.34\\
\cite{LMS}$^c$ & 0.58$\pm$0.09 & 0.45$\pm$0.04 & 0.02$\pm$0.26 &
0.78$\pm$0.12\\
\cite{BKS}$^{c,d}$ & 0.84$\pm$0.12$\pm$0.35 &
0.65$\pm$0.15$\pm^{+0.24}_{-0.23}$ &
0.59$\pm$0.31$\pm^{+0.28}_{-0.25}$ & 1.07$\pm$0.49$\pm$0.35\\
\cite{gatto}$^e$ & 0.79 & 0.55 & 0.28 & 1.01\\
\cite{pardat}$^f$ & 0.80$^{+0.21}_{-0.14}$ & -- & -- & --\\
\end{tabular}
\tablenotes{{}$^a$ QCD Sum Rules}
\tablenotes{{}$^b$ Quark Model}
\tablenotes{{}$^c$ Lattice Calculation}
\tablenotes{{}$^d$ First error statistical, second systematical.}
\tablenotes{{}$^e$ HQET + chiral perturbation theory; value of
$f_+$ taken
{}from experiment.}
\tablenotes{{}$^f$ Experiment (using pole dominance)}
\end{table}

\begin{table}
\caption{The formfactors of the $b\to u$ transitions at $t=0$ in
different models.}
\begin{tabular}{lllll}
Reference & $f_+^{B\to \pi}$ & $A_1^{B\to \rho}$ & $A_2^{B\to
\rho}$ & $V^{B\to \rho}$ \\ \tableline
This paper & 0.26$\pm$0.02 & 0.5$\pm$0.1 & 0.4$\pm$0.2 &
0.6$\pm$0.2\\
\cite{ovchi}$^{a,b}$ & 0.26$\pm$0.01 & -- & -- & --\\
\cite{slobo}$^{a,b}$ & -- & 0.96$\pm$0.15 & 1.21$\pm$0.18 & 1.27$
\pm$0.12\\
\cite{narrdecay}$^a$ & 0.23$\pm$0.02 & 0.35$\pm$0.16 &
0.42$\pm$0.12 &
0.47$\pm$0.14 \\
\cite{dompfaffB}$^a$ & 0.4$\pm$0.1 & -- & -- & -- \\
\cite{BWS}$^c$ & 0.33 & 0.28 & 0.28 & 0.33\\
\cite{GISW}$^c$ & 0.09 & 0.05 & 0.02 & 0.27\\
\cite{gatto}$^d$ & 0.89 & 0.21 & 0.20 & 1.04\\
\end{tabular}
\tablenotes{{}$^a$ QCD Sum Rules}
\tablenotes{{}$^b$ Analysis suffering from a missing factor 12 in the
perturbative contribution.}
\tablenotes{{}$^c$ Quark Model}
\tablenotes{{}$^d$ HQET + chiral perturbation theory.}
\end{table}

\begin{table}
\caption{Decay rates of the $c\to d$ transitions in units
$|V_{cd}|^2\, 10^{11}\,{\rm s}^{-1}$. $\Gamma_L$ denotes
the portion of the rate with a longitudinal polarized $\rho$,
$\Gamma_T$ with a transversely polarized $\rho$, $\Gamma_+$ with a
$\rho$ with positive, $\Gamma_-$ with a $\rho$ with negative
helicity.}
\begin{tabular}{lllll}
Reference & $\Gamma(D^0\to\pi^-e^+\nu)$ & $\Gamma(D^0\to\rho^-e^+
\nu)$ & $\Gamma_L/\Gamma_T$ & $\Gamma_+/\Gamma_-$\\ \tableline
This paper & 0.80$\pm$0.17 & 0.24$\pm$0.07 & 1.31$\pm$0.11
& 0.24$\pm$0.03\\
\cite{unpub} & 1.45$^{+0.95}_{-0.71}$ & -- & -- & -- \\
\cite{dompfaffD}$^a$ & 1.66$^{+0.23}_{-0.21}$ & -- & -- & -- \\
\cite{slobo}$^b$ & -- & 1.4$\pm$1.0 & 0.9 & -- \\
\cite{BWS}$^a$ & 1.41 & 1.38 & 0.91 & 0.19\\
\cite{KS}$^c$ & 1.41 & 1.40 & 0.80 & 0.13\\
\cite{GISW} & 0.77 & 1.35 & 1.33 & 0.11\\
\cite{LMS}$^{a,d}$ & 0.99$^{+0.34}_{-0.28}$ & 0.83$\pm$0.19 &
1.86$\pm$0.56 & 0.16\\
\cite{BKS}$^{a,d}$ & 2.09$^{+2.24}_{-1.44}$ & 1.09 & 1.10 & 0.18\\
\cite{gatto}$^{a,e}$ & 1.9 & 0.93 & 1.40 & 0.14\\
\cite{pardat}$^f$ & $\left|0.22/V_{cd}
\right|^2\cdot(1.9^{+1.1}_{-0.6})$
& \multicolumn{3}{l}{$1/|V_{cd}|^2\,\Gamma(D^+\to\rho^0 e^+\nu) <
0.71\cdot 10^{11}\,{\rm s}^{-1}$}\\
\end{tabular}
\tablenotes{{}$^a$ Rates calculated using pole dominance with
$m_{1^-}
= 2.01\,$GeV, $m_{1^+} = 2.42\,$GeV.}
\tablenotes{{}$^b$ No values of formfactors given.}
\tablenotes{{}$^c$ Values of formfactors at $t=0$ identical to
\cite{BWS}.}
\tablenotes{{}$^d$ Values without errors from central values of
Table~I.}
\tablenotes{{}$^e$ Value of $\Gamma(D^0\to\pi^-e^+\nu)$ taken from
experiment.}
\tablenotes{{}$^f$ Experiment}
\end{table}

\begin{table}
\caption{Decay rates of the $b\to u$ transitions in units
$|V_{ub}|^2\, 10^{13}\,{\rm s}^{-1}$. $\Gamma_L$ denotes
the portion of the rate with a longitudinal polarized $\rho$,
$\Gamma_T$ with a transversely polarized $\rho$, $\Gamma_+$ with a
$\rho$ with positive, $\Gamma_-$ with a $\rho$ with negative
helicity.}
\begin{tabular}{lllll}
Reference & $\Gamma(\bar B^0\to\pi^+e^-\bar\nu)$ &
$\Gamma(\bar B^0\to\rho^+e^-
\bar\nu)$ & $\Gamma_L/\Gamma_T$ & $\Gamma_+/\Gamma_-$\\ \tableline
This paper & 0.51$\pm$0.11 & 1.2$\pm$0.4 & 0.06$\pm$0.02 &
0.007$\pm$0.004\\
\cite{ovchi} & 0.68$\pm$0.23 & -- & -- & --\\
\cite{slobo} & -- & 0.77$\pm$0.42 & -- & --\\
\cite{narrdecay}$^a$ & 0.302$\pm$ 0.005 & 3.3$\pm$0.3 &
0.88$^{+0.39}_{-0.20}$ & 0.12$^{+0.04}_{-0.02}$\\
\cite{dompfaffB}$^b$ & 1.45$\pm$0.59& -- & -- & --\\
\cite{BWS}$^a$ & 0.74 & 2.6 & 1.34 & 0.16\\
\cite{KS}$^c$ & 0.74 & 2.30 & 0.54 & 0.02\\
\cite{GISW} & 0.21 & 1.63 & 0.75 & 0.08\\
\cite{gatto}$^a$ & 5.4 & 3.4 & 0.36 & 0.14\\
\end{tabular}
\tablenotes{{}$^a$ Rates calculated using pole dominance with
$m_{1^-} = 5.33\,$GeV, $m_{1^+} = 5.71\,$GeV.}
\tablenotes{{}$^b$ Rate calculated using a modified pole dominance
with
$m_{1^-} = 5.33\,$GeV.}
\tablenotes{{}$^c$ Values of formfactors at $t=0$ identical to
\cite{BWS}.}
\end{table}

\figure{The integration region for the perturbative contributions to
the sum rules in the $s_L$-$s_H$ plane (without non-Landau
contributions).\label{fig:intgebiet}}

\figure{(a) The formfactor $f_+^{D\to\pi}(0)$ as function of the
Borel
parameter $M_c^2$. The parameter sets in the legend are $m_c =
1.3\,{\rm GeV}$, $s_D^0=6\,{\rm GeV}^2$, $s_\pi^0 = 0.75\,
{\rm GeV}^2$
(set C1), $m_c = 1.3\,{\rm GeV}$, $s_D^0=6\,{\rm GeV}^2$, $s_\pi^0 =
1\,{\rm GeV}^2$ (set C2), $m_c = 1.4\,{\rm GeV}$, $s_D^0=7\,{\rm
GeV}^2$, $s_\pi^0 = 0.75\,{\rm GeV}^2$ (set C3), $m_c = 1.4\,{\rm
GeV}$, $s_D^0=7\,{\rm GeV}^2$, $s_\pi^0 = 1\,{\rm GeV}^2$ (set C4).
(b)
$f_+^{B\to\pi}(0)$ as function of the Borel parameter $M_b^2$ with
the
parameter sets B1 ($m_b = 4.6\,{\rm GeV}$, $s_B^0=36\,{\rm GeV}^2$,
$s_\pi^0 = 0.75\,{\rm GeV}^2$), B2 ($m_b = 4.8\,{\rm GeV}$,
$s_B^0=36\,{\rm GeV}^2$, $s_\pi^0 = 1\,{\rm GeV}^2$), B3
($m_b = 4.8\,
{\rm GeV}$, $s_B^0=34\, {\rm GeV}^2$, $s_\pi^0 = 0.75\,{\rm GeV}^2$),
B4 ($m_b = 4.8\,{\rm GeV}$, $s_B^0=34\,{\rm GeV}^2$, $s_\pi^0 =  1\,
{\rm GeV}^2$).\label{fig:fplus0}}

\figure{(a) The formfactor $f_+^{D\to\pi}(t)/f_+^{D\to\pi}(0)$ as
function of $t$ for all parameter sets at $M_c^2=3\,\mbox{GeV}^2$.
A pole fit yields $m_{pol}=(1.95\pm 0.10)\,{\rm GeV}$, the maximum
physical value of $t$ is $t^{max} = 2.96\,\mbox{GeV}^2$. (b) Like
(a) for
$f_+^{B\to\pi}(t)/f_+^{B\to\pi}(0)$. $m_{pol} = (5.25\pm 0.10)\,{\rm
GeV}$ (for B3 and B4), $t^{max} = 26.4\,
\mbox{GeV}^2$.\label{fig:fplust}}

\figure{The electron spectra $d\Gamma/dE$ as function of the electron
energy $E$ for (a) the decay $D\to \pi e \nu$ (set C1, $M_c^2 = 3\,
{\rm GeV}^2$), (b) $B\to \pi e \nu$ (set B3, $M_b^2 = 8\,{\rm
GeV}^2$).\label{fig:specpi}}

\figure{The formfactors of $D\to\rho$ at $t=0$ as functions of the
Borelparameter $M_c^2$. The parameter sets are C5 ($m_c = 1.3\,{\rm
GeV}$, $s_D^0=6\,{\rm GeV}^2$, $s_\rho^0 = 1.25\,{\rm
GeV}^2$), C6 ($m_c = 1.3\,{\rm GeV}$, $s_D^0=6\,{\rm GeV}^2$,
$s_\rho^0 = 1.5\,{\rm GeV}^2$), C7 ($m_c = 1.4\,{\rm GeV}$,
$s_D^0=7\,
{\rm GeV}^2$, $s_\rho^0 = 1.25\,{\rm GeV}^2$), C8 ($m_c = 1.4\,{\rm
GeV}$, $s_D^0=7\,{\rm GeV}^2$, $s_\rho^0 = 1.5\,{\rm GeV}^2$).
(a) $A_1^{D\to\rho}(0)$, (b)
$A_2^{D\to\rho}(0)$, (c) $V^{D\to\rho}(0)$.\label{fig:FF0D}}

\figure{The formfactors of $B\to\rho$ at $t=0$ as functions of the
Borelparameter $M_b^2$. The parameter sets are B5 ($m_b =
4.6\,{\rm GeV}$, $s_B^0=36\,{\rm GeV}^2$, $s_\rho^0 = 1.25\,{\rm
GeV}^2$), B6 ($m_b = 4.6\,{\rm GeV}$, $s_B^0=36\,{\rm GeV}^2$,
$s_\rho^0 = 1.5\,{\rm GeV}^2$), B7 ($m_b = 4.8\,{\rm GeV}$,
$s_B^0=34\,
{\rm GeV}^2$, $s_\rho^0 = 1.25\,{\rm GeV}^2$), B8 ($m_b = 4.8\,{\rm
GeV}$, $s_B^0=34\,{\rm GeV}^2$, $s_\rho^0 = 1.5\,{\rm GeV}^2$).
 (a) $A_1^{B\to\rho}(0)$, (b)
$A_2^{B\to\rho}(0)$, (c) $V^{B\to\rho}(0)$.\label{fig:FF0B}}

\figure{The formfactors of $D\to\rho$, normalized to their values at
$t=0$, as functions of $t$ for all parameter sets and
$M_c^2 = 3\,{\rm
GeV}^2$. A pole-fit is sensible only for the vector formfactor and
yields $m_{pol} = (2.5\pm 0.2)\,{\rm GeV}$. $t_{max} = 1.21\,{\rm
GeV}^2$. (a) $A_1^{D\to\rho}(t)/A_1^{D\to\rho}(0)$, (b)
$A_2^{D\to\rho}(t)/A_2^{D\to\rho}(0)$,
(c) $V^{D\to\rho}(t)/V^{D\to\rho}(0)$.\label{fig:FFtD}}

\figure{The formfactors of $B\to\rho$, normalized to their values at
$t=0$, as functions of $t$ for all parameter sets and
$M_b^2 = 8\,{\rm
GeV}^2$. A pole-fit is sensible only for the vector formfactor and
yields $m_{pol} = (6.6\pm 0.6)\,{\rm GeV}$. $t_{max} = 20.3\,{\rm
GeV}^2$. (a) $A_1^{B\to\rho}(t)/A_1^{B\to\rho}(0)$, (b)
$A_2^{B\to\rho}(t)/A_2^{B\to\rho}(0)$,
(c) $V^{B\to\rho}(t)/V^{B\to\rho}(0)$.\label{fig:FFtB}}

\figure{The electron spectrum $d\Gamma/dE$ for $D\to \rho e \nu$ as
function of the electron energy $E$ (set C5, $M_c^2 = 3\,
{\rm GeV}^2$).
\label{fig:specDrho}}

\figure{(a) The electron spectrum $d\Gamma/dE$ for $B\to \rho e
\nu$ as
function of the electron energy $E$ (set B8, $M_b^2 = 8\,{\rm
GeV}^2$). (b) Comparison of the spectra $1/\Gamma\, d\Gamma/dE$ as
functions of $E$ as obtained in this paper (parameters like (a)) and
in the BWS \cite{BWS} and the GISW model \cite{GISW}. The chosen
normalization emphasizes the shape of the
spectra.\label{fig:specBrho}}

\figure{Comparison of the spectra $d\Gamma/dE$ as functions of E
of the exclusive decays $B\to
\pi,\,\rho e \nu$ (parameters like in Figs.~\ref{fig:specpi} and
\ref{fig:specBrho}(a)) to the spectrum of the inclusive decay $B\to
X_u e \nu$, taken from \cite{inklupreprint} and calculated with the
same parameters. (a) restsystem of the decaying $B$, (b) labsystem of
a collider working at the $\Upsilon(4S)$
resonance.\label{fig:inklusiv}}

\figure{Diagrams contributing to the Wilson-coefficient of the gluon
condensate. Lines with a cross denote vacuum expectation values. $0$,
$x$, $y$ are space-time coordinates, $b$, $u$, $q$ denote quark
flavours, $q$ being a light quark ($u$ or $d$). The weak vertex is at
$y$.\label{fig:diagrams}}

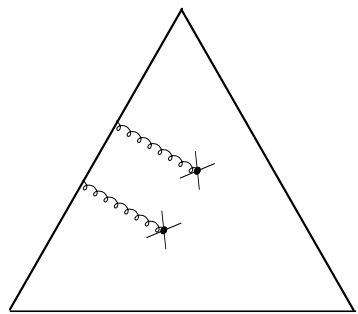
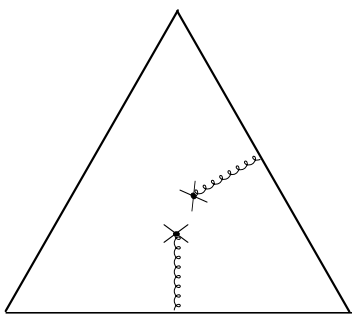
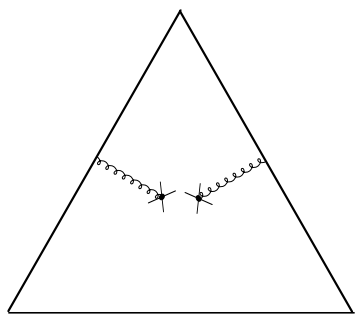
\end{document}